\documentclass[10pt]{llncs}

\usepackage{cite}
\usepackage{mathptmx}
\usepackage{graphicx}
\usepackage{times}
\usepackage{amsmath}
\usepackage{amsfonts}
\usepackage{caption}
\usepackage{multirow}
\usepackage{paralist}
\usepackage{listings}
\usepackage{wrapfig}

\usepackage{color}
\definecolor{RED}{rgb}{1,0,0}
\definecolor{BLUE}{rgb}{0,0,1}

\newcommand{\lref}[1]{Listing~\ref{#1}}
\newcommand{\fig}[1]{Figure~\ref{#1}}

\begin{document}
\lstset{language=[11]C++,basicstyle=\scriptsize,captionpos=b}
%
\title{From Big Data to Big Displays\\
High-Performance Visualization at Blue Brain}

%
%
%
%

\author{Stefan~Eilemann, Marwan~Abdellah, Nicolas~Antille, Ahmet~Bilgili, Grigory~Chevtchenko, Raphael~Dumusc, Cyrille~Favreau, Juan~Hernando, Daniel~Nachbaur, Pawel~Podhajski, Jafet~Villafranca, Felix Sch\"urmann\thanks{\textit{firstname.lastname}@epfl.ch}}
\institute{Blue Brain Project, Ecole Polytechnique Federale de Lausanne}

\maketitle

\begin{abstract}
  Blue Brain has pushed high-performance visualization (HPV) to complement its
  HPC strategy since its inception in 2007. In 2011, this strategy has been
  accelerated to develop innovative visualization solutions through increased
  funding and strategic partnerships with other research institutions.

  We present the key elements of this HPV ecosystem, which integrates C++
  visualization applications with novel collaborative display systems. We
  motivate how our strategy of transforming visualization engines into services
  enables a variety of use cases, not only for the integration with
  high-fidelity displays, but also to build service oriented architectures, to
  link into web applications and to provide remote services to Python
  applications.
\end{abstract}

\section{Motivation}

The Blue Brain Project (BBP) uses simulation-based research to analyze and
reverse engineer cortical neuron circuits. The simulated models go beyond using
detailed models of individual neurons or large-scale network models of
simplified neurons, they model in the order of hundreds of thousands of detailed
neurons in a fully connected circuit. The project generates a multitude of data
for the model building and during the simulation of these models. This data
ranges from terabyte-sized image stacks for data extraction to detailed
in-silico circuit models of large geometric complexity and terabyte-size
simulation reports.

Visualization supports the BBP along all parts of the project to understand and
debug model data, building and simulation algorithms as well as validating and
discovering new insight from the in-silico experiments. Our strategy to support
this mission is based on components linked through network protocols:
High-fidelity display systems to see more detail in complex data, a set of
standard rendering engines (rasterization, out-of-core volume rendering,
interactive raytracing), and decoupled, light-weight applications using
these components remotely. In the following we will present these components
along with a few use cases.

\section{High-Fidelity Displays}

High-fidelity display systems are the integration point of the Blue Brain
visualization capabilities. They are the evolution of existing visualization
systems, enabling high resolution, immersion and team collaboration. Compared
to current single-user or single-presenter systems, collaborative display
systems allow real team work through a combination of size, resolution and user
friendly implementation. Compared to immersive visualization systems like the
CAVE, they provide a more approachable environment for high-fidelity
visualization. For all use cases, the increased display size and resolution
allows better data exploration for 2D and 3D content, facilitating large data
exploration.

\subsection{Tiled Multitouch Display Walls}

\begin{figure}[h!t]
  \includegraphics[width=\columnwidth]{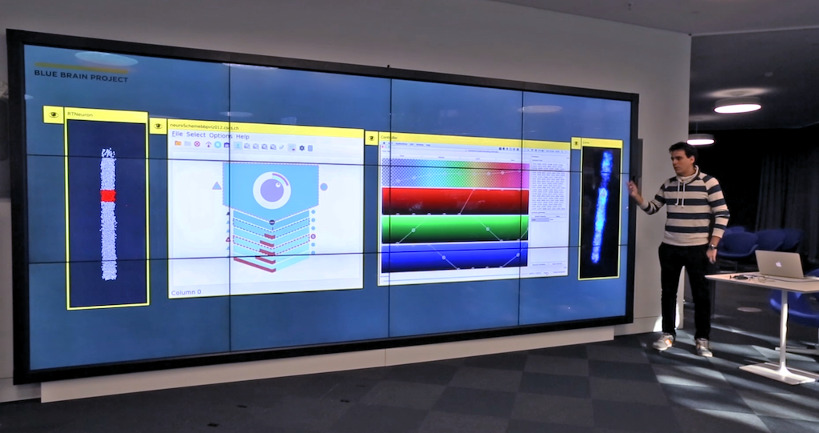}
  \caption{\label{fTDW}Blue Brain $4\times 3$ tiled display wall}
\end{figure}

The core of the Blue Brain visualization infrastructure are multiple
high-resolution tiled display walls driven by our Tide software~\cite{tide}. All
walls are equipped with a multitouch user interface and can be remote controlled
from any web browser. The walls are built using thin-bezel, 55 inch, Full HD LCD
panels with a hardened glass sheet. We use $4\times 3$ and $3\times 3$
configurations for a total of 24 and 18 Megapixel resolution, respectively. The
display size of over five meter diagonally (four meter for $3\times 3$) allows
team-size collaboration (up to ten people) or project-wide presentations (up to
a hundred people). \fig{fTDW} shows one wall during a project-wide presentation
with multiple interactive applications running remotely on the wall.

\subsection{Tide}

\begin{figure}[h!t]
  \includegraphics[width=\columnwidth]{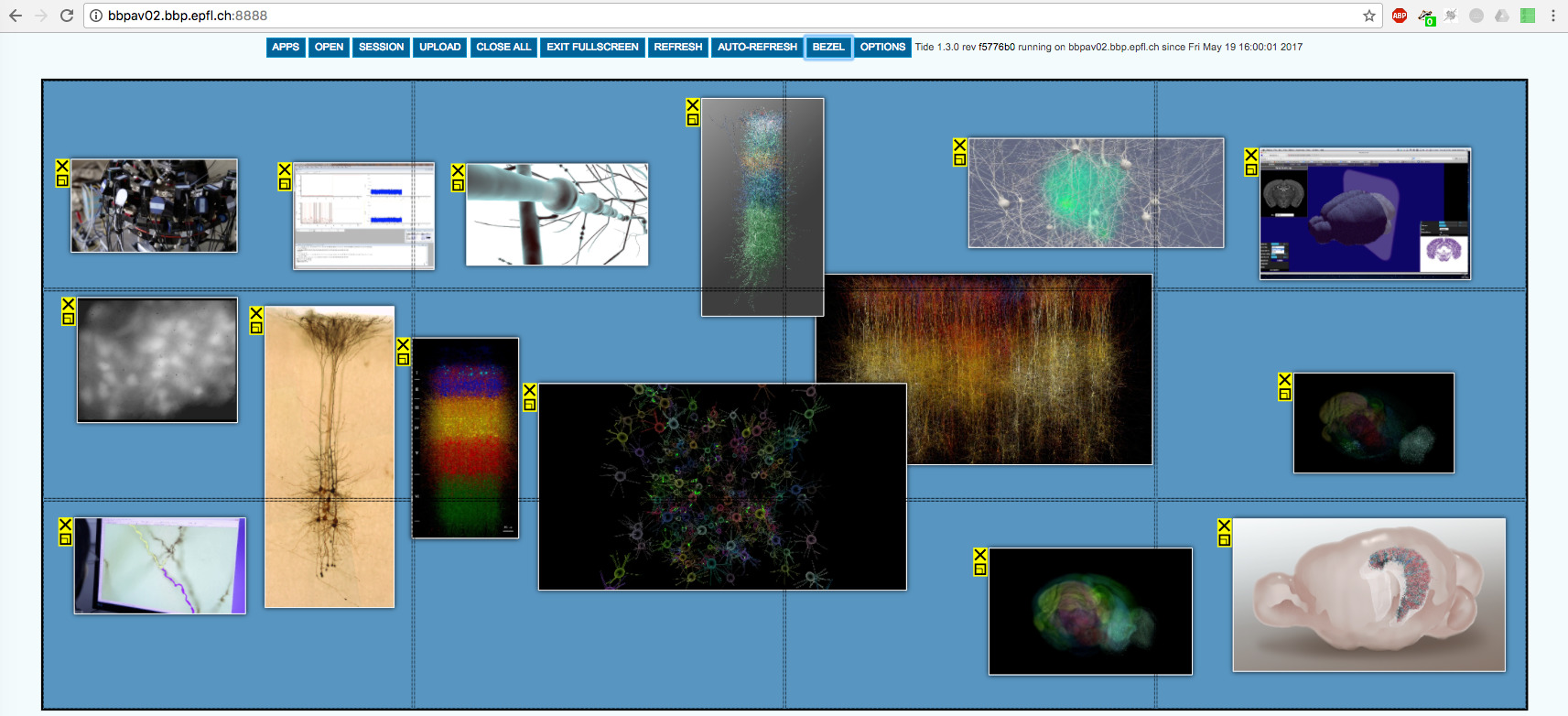}
  \caption{\label{fTideWeb}Tide web interface}
\end{figure}

Tide (Tiled Interactive Display Environment) is the software driving the Blue
Brain tiled display walls. It provides multi-window, multi-user touch
interaction on large surfaces --- think of a giant collaborative wall-mounted
tablet. Tide is a distributed application that can run on multiple machines to
power display walls or projection systems of any size. Its user interface is
designed to offer an intuitive experience on touch walls. It works just as
well on non touch-capable installations by using its web interface from any
web browser. \fig{fTDW} shows Tide on a $4\times 3$ display wall and
\fig{fTideWeb} shows the Tide web interface in a browser.

While there is substantial research on tiled display wall software~\cite{Sage,
Sage2, Omegalib, DK:11, JLMV:06, DisplayCluster}, we found that most solutions
were not ready for production use in a $24\times 7$ unattended environment. For
this reason, we started with the TACC DisplayCluster open source
software~\cite{DisplayCluster}, which we incrementally refactored and improved
to the current TIDE implementation. On the other hand, the hardware has been
commoditized to make these type of installations affordable to medium-sized
institutions which allowed us to build the software integration for a reasonable
startup cost. 
We have focused on the multitouch user interface, which implements
a low entry barrier for new users, a unique capability of our solution.

Tide supports three types of content: files (high-resolution images, movies,
pdfs), built-in applications (web browser, whiteboard) and remote applications
using the Deflect library (DesktopStreamer, Equalizer-based applications,
Brayns). The multitouch user interface can handle multiple users manipulating
different windows and their content simultaneously.

\subsection{Deflect}

Deflect is the client library for Tide. It provides an API for pixel streaming
to Tide and for receiving events from Tide. The pixel streaming allows
synchronized parallel streaming from parallel rendering application as well as
monoscopic and stereoscopic streams. Various events allow the application to
react to multi-touch input from the wall.

Deflect is integrated into the Equalizer parallel rendering
framework~\cite{EMP:09}, enabling transparent usage of Equalizer applications on
Tide walls. Furthermore, the DesktopStreamer application mirrors the desktop of
other machines onto a wall window and allows interaction with the remote
desktop. Other rendering applications, such as our interactive raytracing
engine Brayns~\cite{brayns} are easily integrated with Deflect and Tide.

\subsection{OpenDeck}

\begin{figure}[h!t]
  \includegraphics[width=\columnwidth]{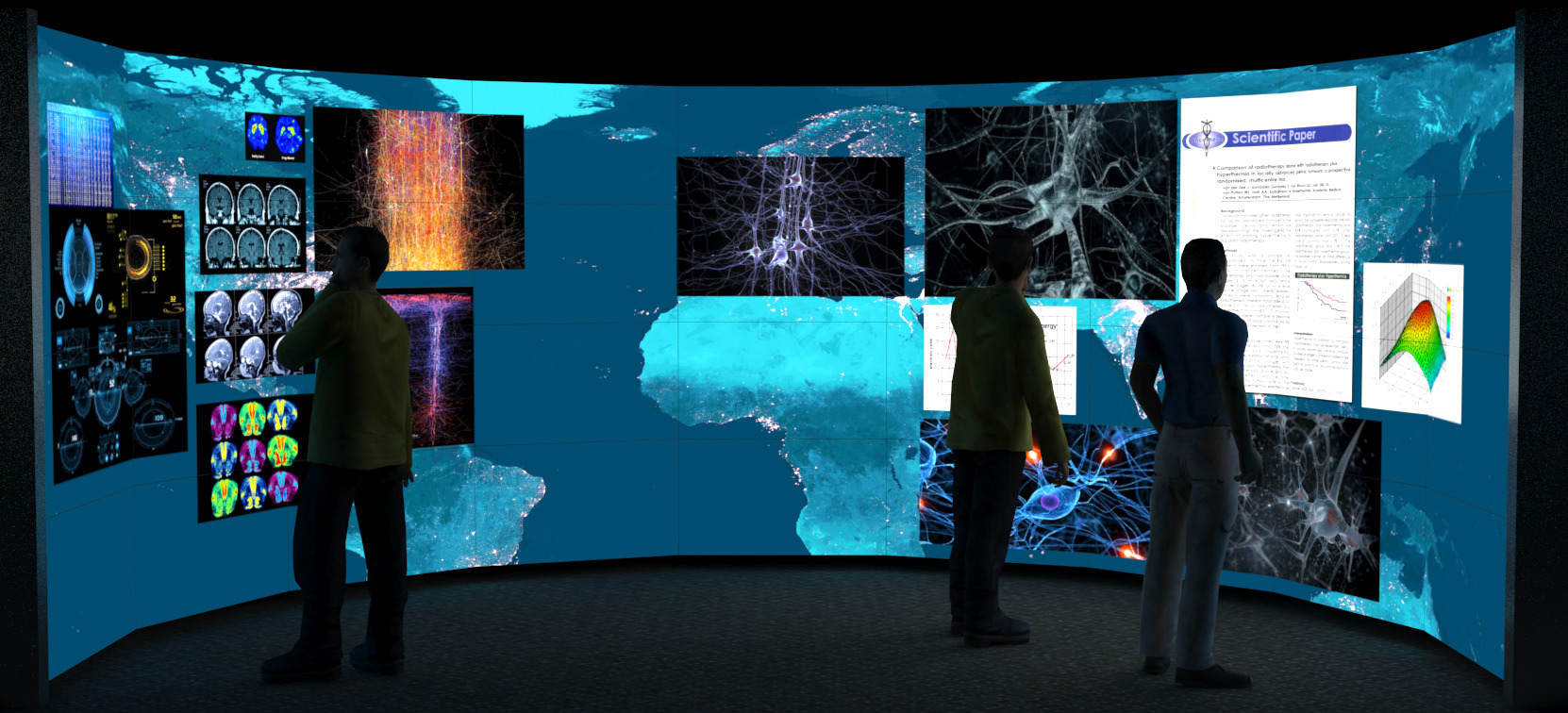}
  \caption{\label{fOpenDeck}OpenDeck concept rendering}
\end{figure}

OpenDeck is our next-generation visualization system, aiming to integrate the
success of tiled display walls with a seamless transition to fully immersive
environments. We are currently in the process of installing a system which
consists of a semi-cylindrical back-projection screen with 41 Megapixel usable
resolution on a 36 $m^2$ surface (\fig{fOpenDeck}). Like the display walls, it
is equipped with multitouch capabilities which makes it usable as a monoscopic
collaboration system from the first day of installation. Unlike tiled display
walls, it is active stereo capable and equipped with a 3D tracking system for
immersive rendering. For increased immersion, a lower resolution front
projection system fills in the floor area. OpenDeck will run TIDE and our
immersive applications based on Equalizer~\cite{EMP:09} once the system is
installed.

OpenDeck will provide a unique environment for the research and development of
new visualization techniques. It will open a set of questions along immersive
touch user interfaces, transitions and mixing of monoscopic to immersive usage,
the combination of tracked and touch devices, multi-user immersion, latency
reduction for remote immersive rendering as well as multi-site collaboration.

\section{Rendering Applications}

The rendering applications form the backbone of our ecosystem. They cover a
wide range of established rendering algorithms to serve a broad set of use
cases for visual debugging, scientific illustrations and communication.

\subsection{Brayns: Interactive Raytracing}

\begin{figure}[ht]\center
  \includegraphics[height=5cm]{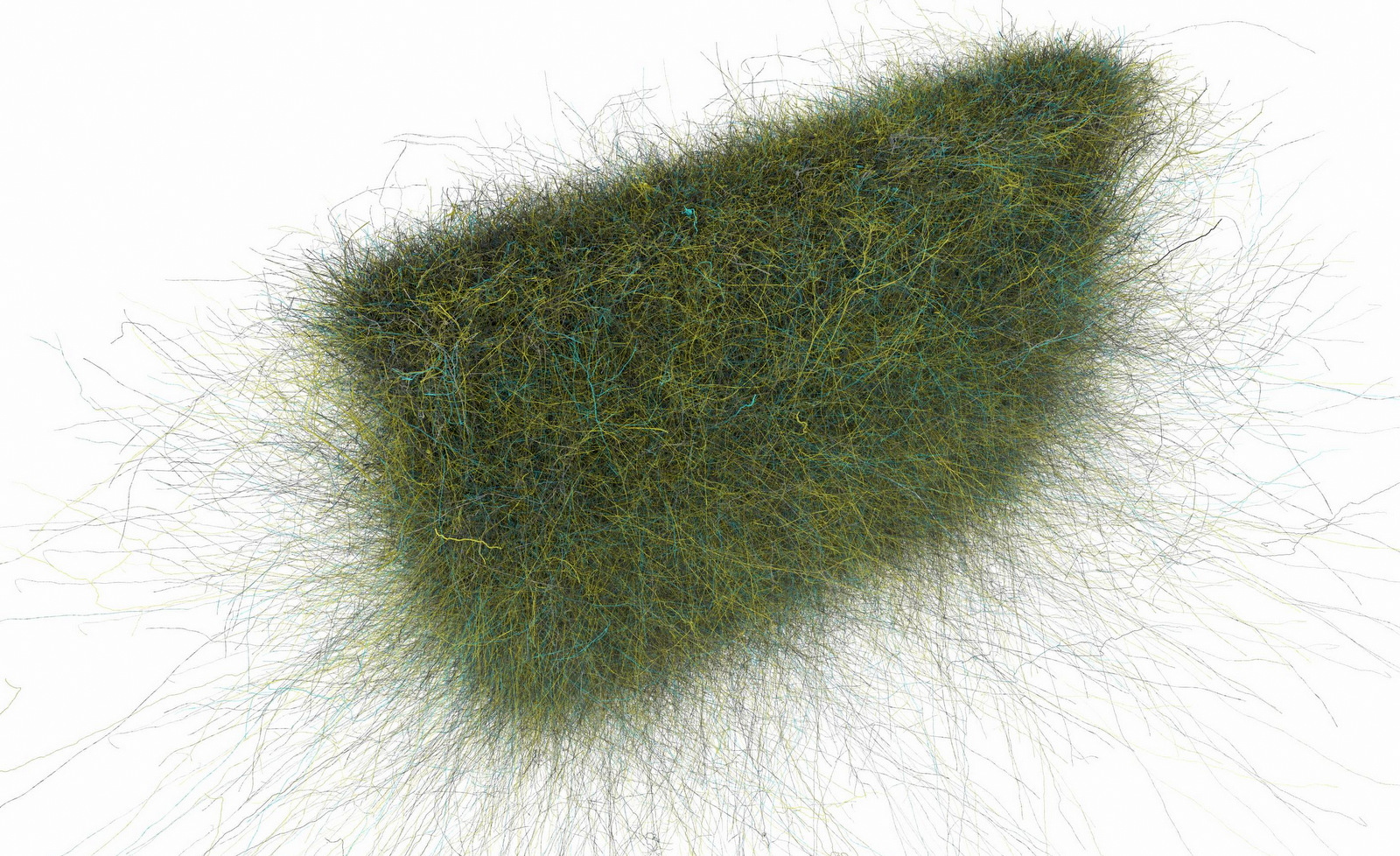}\hfil
  \includegraphics[height=5cm]{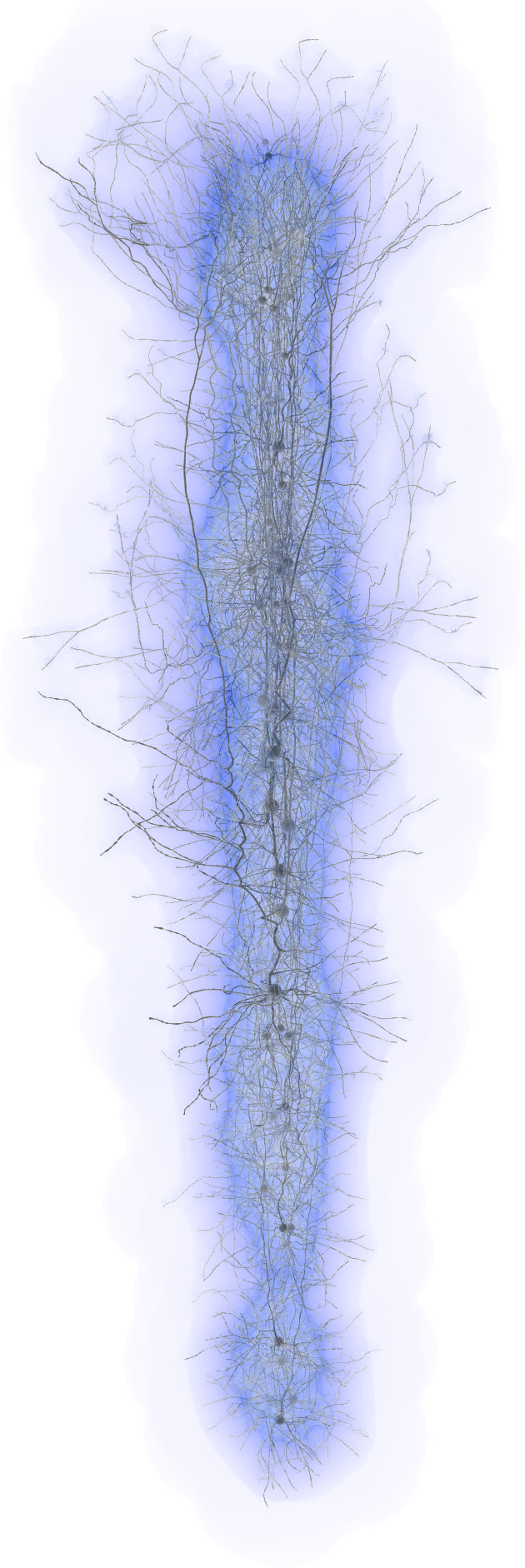}
  \caption{\label{fBrayns}Interactive raytracing for neuroscience}
\end{figure}

Advances in computer hardware and software have brought raytracing to the point
where it replaces classical rasterization for virtually all use cases in
scientific visualization. On one hand, OpenGL-like local illumination for
typical data sets used with rasterization (up to hundred million triangles) can
be done at similar framerates to OpenGL~\cite{ospray}. For small datasets,
OpenGL performs better, but for larger data sets raytracing outperforms OpenGL.
This is due to a better scalability with respect to the data set size
($O(\log{N})$ vs $O(n)$). Furthermore, CPU-based raytracing allows the rendering
of larger data sets without any level-of-detail algorithms due to the larger
memory size. Last, but not least, advanced rendering algorithms such as
shadowing, reflections and global illumination (\fig{fBrayns}, left) are
significantly easier to implement in a raytracer. The only area where
rasterization provides a benefit is for rendering at very high frame rates,
needed for example for immersive visualization, or very high resolutions.

It is for these reasons that interactive raytracing is the future technology for
interactive and offline rendering at Blue Brain. We have developed a first open
source implementation of a visualization engine with different backends:
OSPRay~\cite{ospray} for CPU-based raytracing and OptiX~\cite{optix} for
GPU-based raytracing. This application called Brayns can load and visualize our
data sets in a variety of modes and integrates with our messaging solution to be
accessible from Python. We are currently integrating the key algorithms from
Livre for out-of-core volume rendering in Brayns, which will also facilitate
mixing polygonal with large volumetric data in a single scene (\fig{fBrayns},
right).

\subsection{RTNeuron: OpenGL Parallel Rendering}

RTNeuron\cite{HBBES:13} is the oldest of our interactive visualization tools.
Originally conceived as a standalone application for visualizing simulation
results, it has evolved into a rasterization-based rendering engine library
implemented in C++ with a Python wrapping. The power of RTNeuron lies in a
domain specific API designed for the visualization of detailed neuronal circuit
models, this API allows for building custom applications tailored for specific
use cases. It provides features to visualize static circuit data and some
simulation results. Static data visualization includes different visual
representations for neurons, synapses, selective pruning of neuronal trees,
clipping planes and others. The simulation results that can be displayed are the
spiking activity of the neurons and scalar variables from the cable models
mapped to the neuron surface. The color maps for displaying these data are
highly configurable, allowing to apply different color maps to different cell
subsets.

\begin{figure}[h!t]\center
  \includegraphics[width=.49\columnwidth]{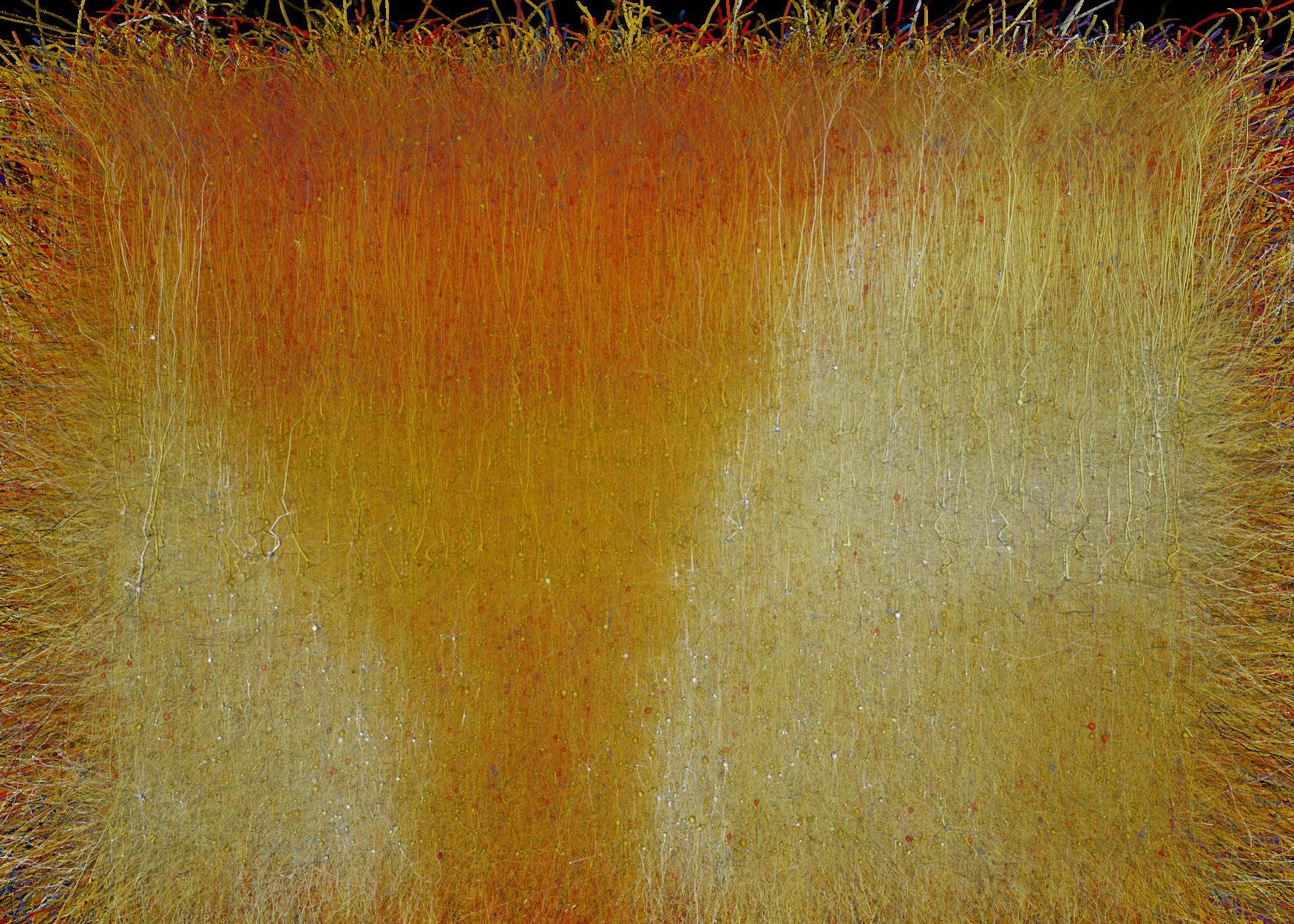}\hfil
  \includegraphics[width=.49\columnwidth]{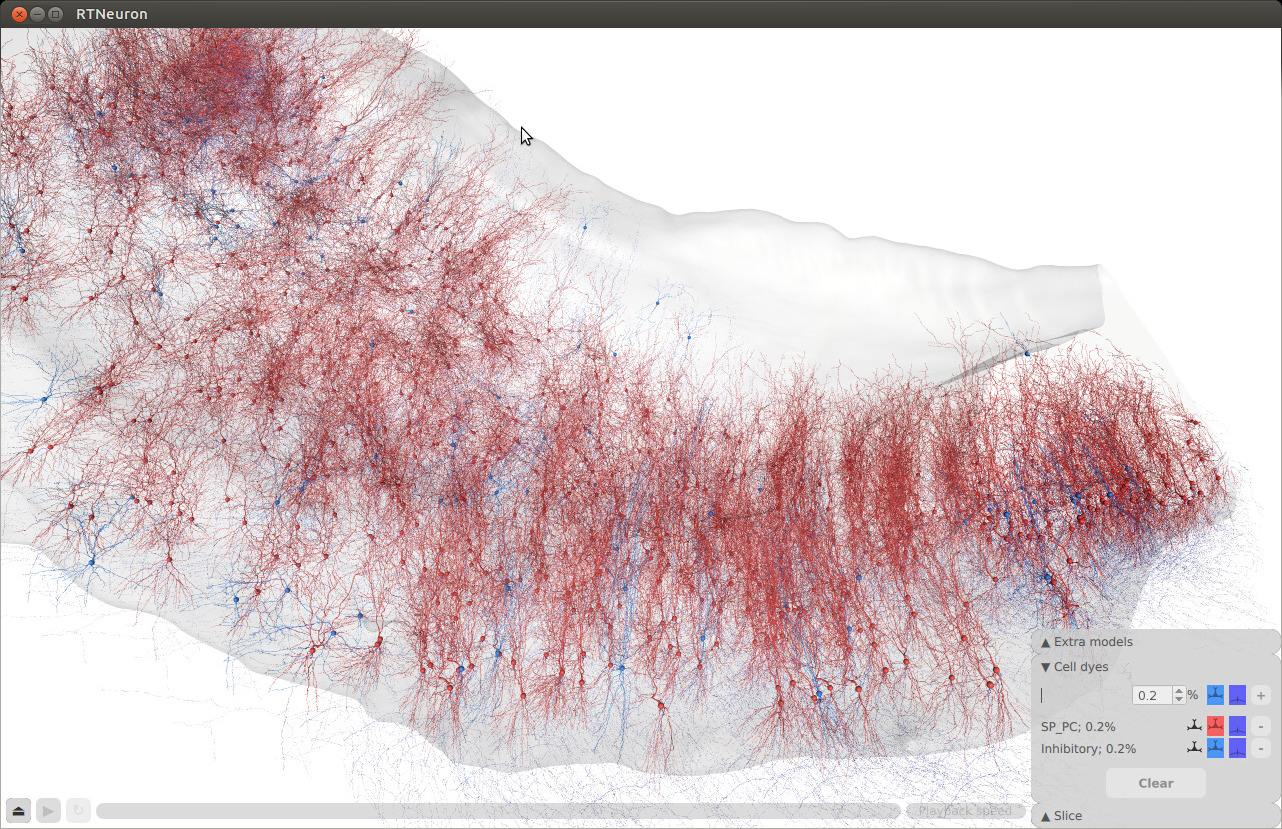}
  \caption{\label{fRTNeuron}RTNeuron rendering the simulated membrame potential of a fraction of the simulation~(left) and a circuit visualization application for hippocampus model validation~(right)}
\end{figure}

Neurons can be displayed with different levels of detail ranging from simple
spheres, cylinder-like geometric models and detailed polygonal meshes. Advanced
rendering techniques included are several types of parallel rendering algorithms
(such as sort-first and sort-last) and several algorithms for transparency
rendering that enable efficient and correct rendering of scenes with great
geometrical complexity and high depth complexity. Transparency is particularly
suitable for masking and highlighting features of interest on the circuit.
Ongoing work is a more scalable parallel rendering algorithm specially suited
for transparency. Tiled rendering is also possible, which allows very high
resolution renderings suitable for printing at sizes larger than A0 (the
original of \fig{fRTNeuron} left is a 36640$\times$26000 pixel image).

The core engine is implemented in C++ using Equalizer and OpenSceneGraph and
leverages part of our messaging framework to allow coupling with other
applications. The 3D view can be embedded inside Qt applications
(\fig{fRTNeuron}, right), in particular in Python with PyQt and QML for
overlaying GUI elements. Several use case specific applications have been built
this way.

\subsection{Livre: Out-of-Core Volume Rendering}

Livre is an interactive volume rendering engine available under a permissive
open source license. Our main contributions are: a state-of-the-art
implementation of an octree-based level-of-detail (LOD) selection, a task
parallel rendering pipeline, a multi-GPU parallel rendering engine, and an
easily extensible renderer through the use of plugin data sources. Our system
brings together state-of-the art algorithms to create a volume rendering engine
capable of handling extremely high-resolution volumes using a high degree of
parallelism, both on a single system and in a distributed cluster. We employ a
GPU-based ray casting algorithm to compute the radiance absorption of the given
volumetric data. The computation is executed per pixel on the pixel shader
hardware of the GPU.

\begin{figure}[t]
    \includegraphics[width=.49\columnwidth]{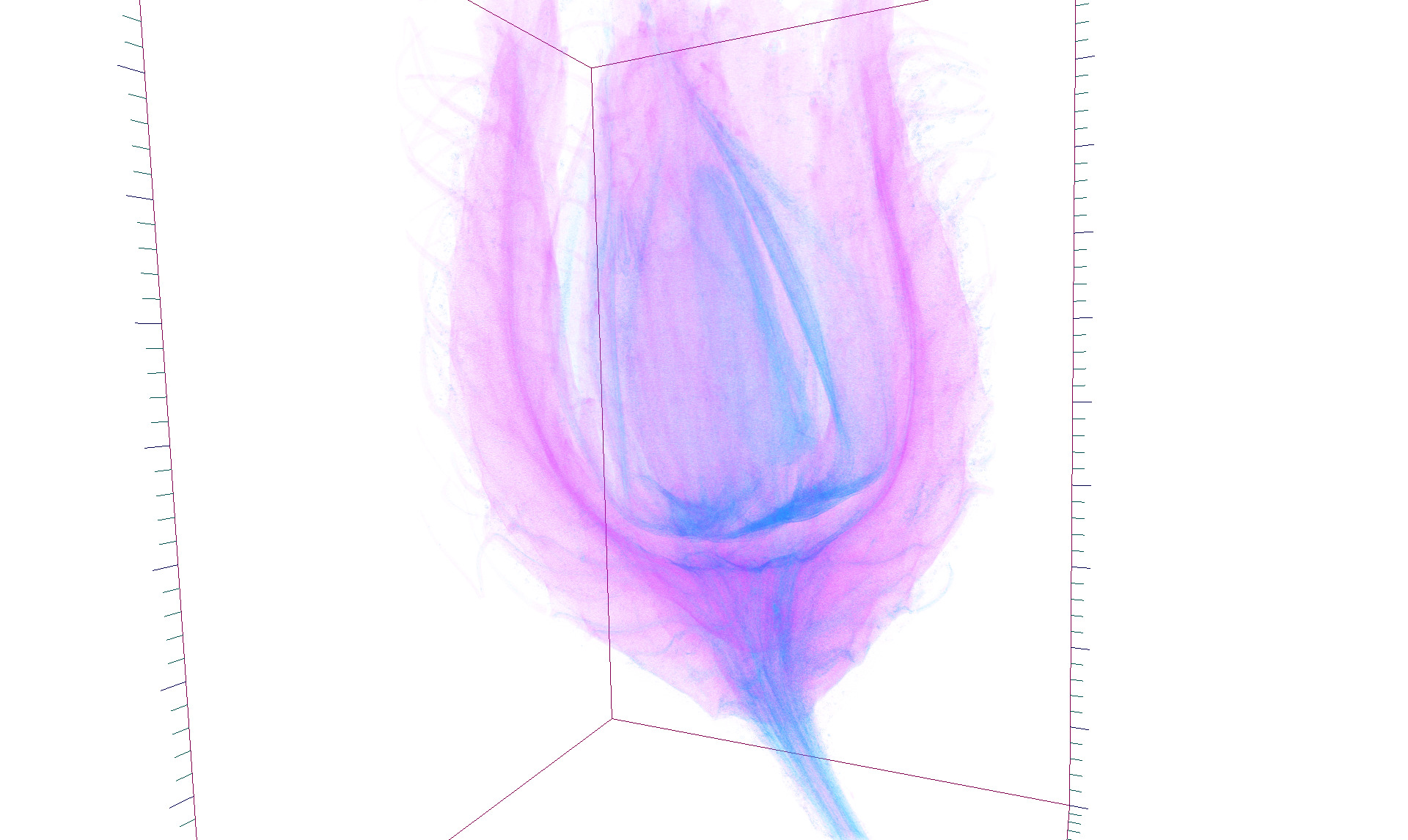}\hfill
    \includegraphics[width=.49\columnwidth]{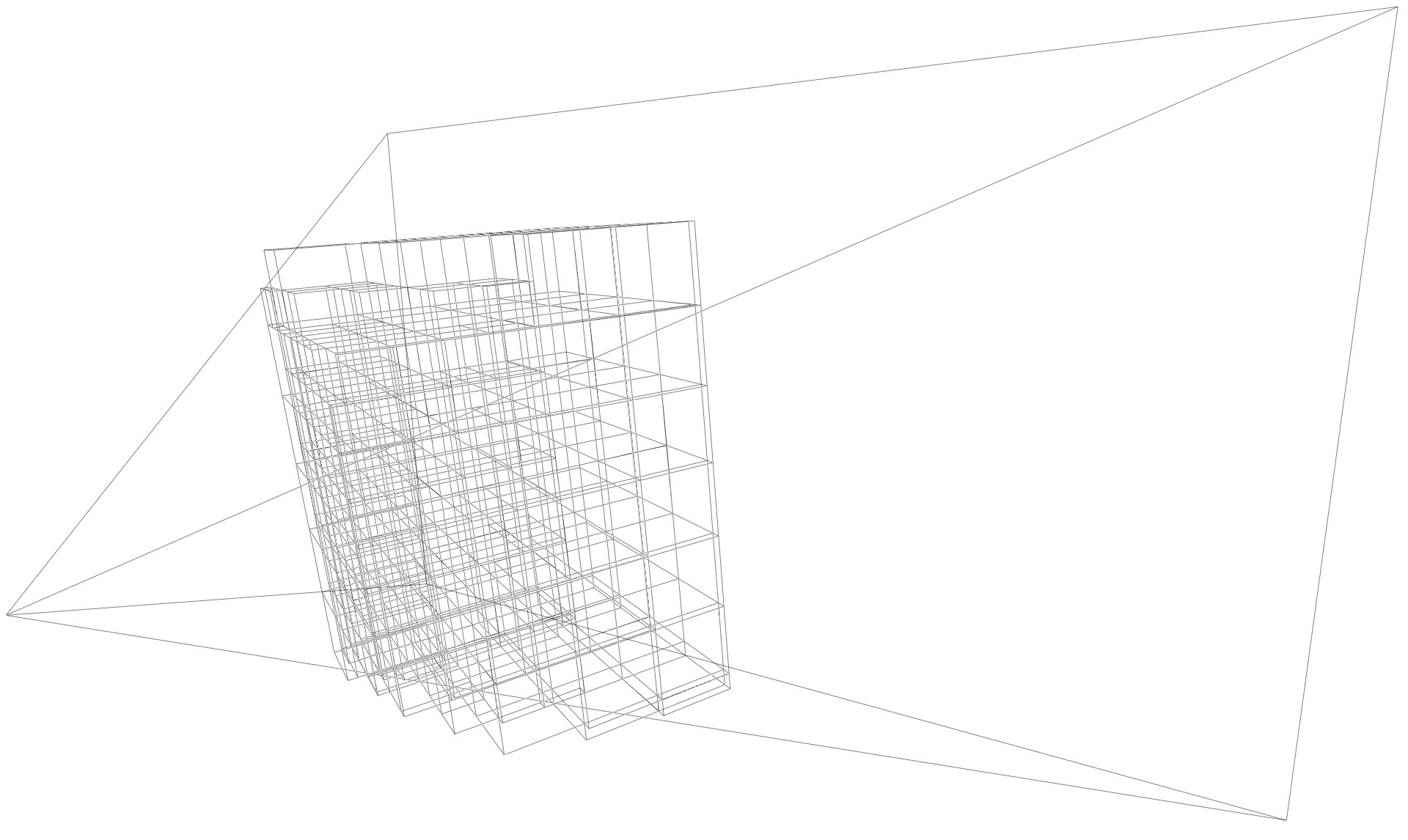}
  \caption{\label{fLivre}Livre rendering a MicroCT dataset (left) and
    the selected LODs (right)}
\end{figure}

In our out-of-core data access layer, multi-resolution data is represented as an
octree data structure. This representation accelerates the selection of the
proper level-of-detail and to track the status of the LODs (in CPU memory, in
GPU memory, not loaded). While rendering, view-based LOD selection is performed
using the screen-space-error (SSE)~\cite{guthe2004} technique. \fig{fLivre}
shows the rendering of a MicroCT dataset with an illustration of the selected
LOD levels.

The creation of volume bricks, their upload to the GPU and the rendering are
executed in separate tasks. These tasks run asynchronously, that is, rendering
is decoupled from data loading. Livre uses a plugin mechanism to access the
volume data, where data sources are implemented as shared libraries and are
loaded on application startup based on the URI of the input data. Data sources
only have to provide the requested volume bricks, that is, there is no defined
file format or even requirement to read the input data from a file system. This
flexibility of the plugin approach lead to novel volume rendering use cases,
where volume representations are created on the fly from different input data
sets, for example from simulation data.

\clearpage
\section{Messaging and Service Architecture}

\begin{wrapfigure}{r}{.382\textwidth}\center\vspace{-6ex}
  \includegraphics[width=.382\textwidth]{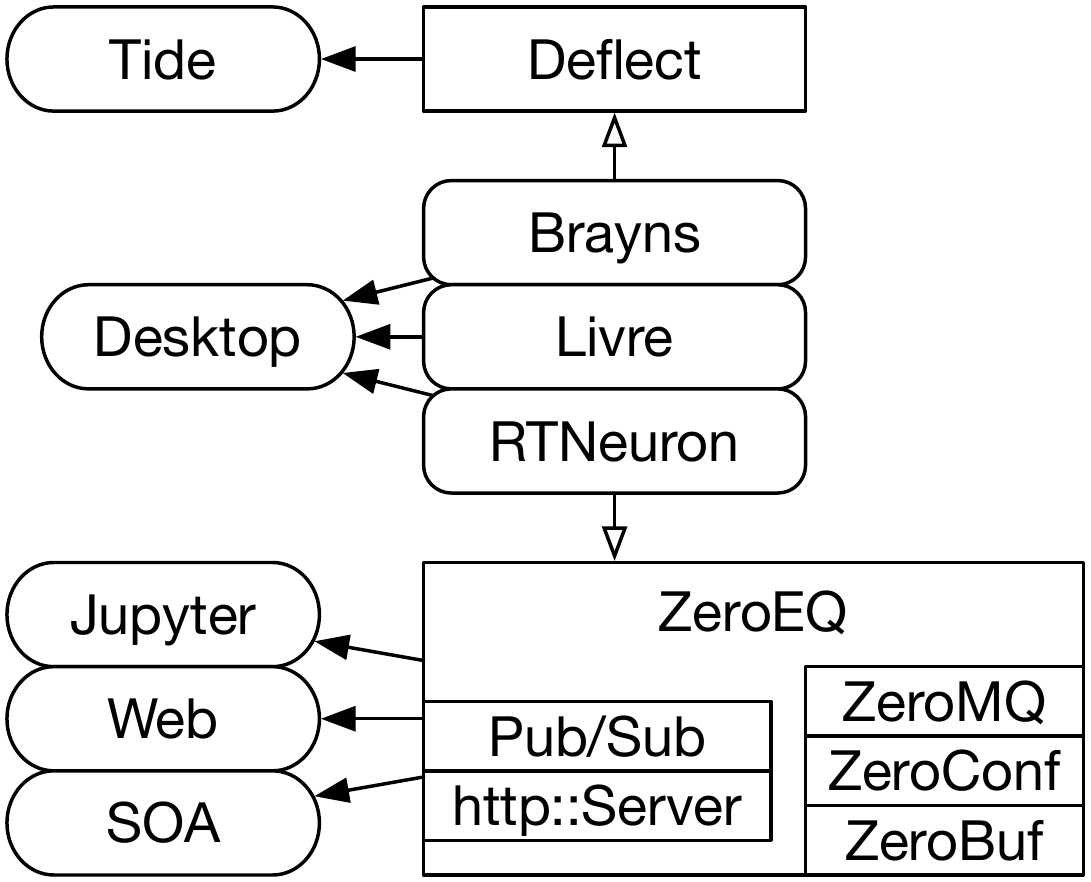}
  {\caption{\label{fZero}Messaging-enabled use cases}}\vspace{-1em}
\end{wrapfigure}

All Blue Brain applications integrate messaging libraries which allow them to be
used as services in a variety of use cases. For example, the Tide web server
providing the user interface shown in \fig{fTideWeb}, is based on this messaging
solution. Other use cases are remote python APIs, JavaScript user interfaces and
service architectures combining multiple visualization applications with data
providers such as HPC simulations (\fig{fZero}).

\begin{figure}[ht]\center
  \includegraphics[width=\columnwidth]{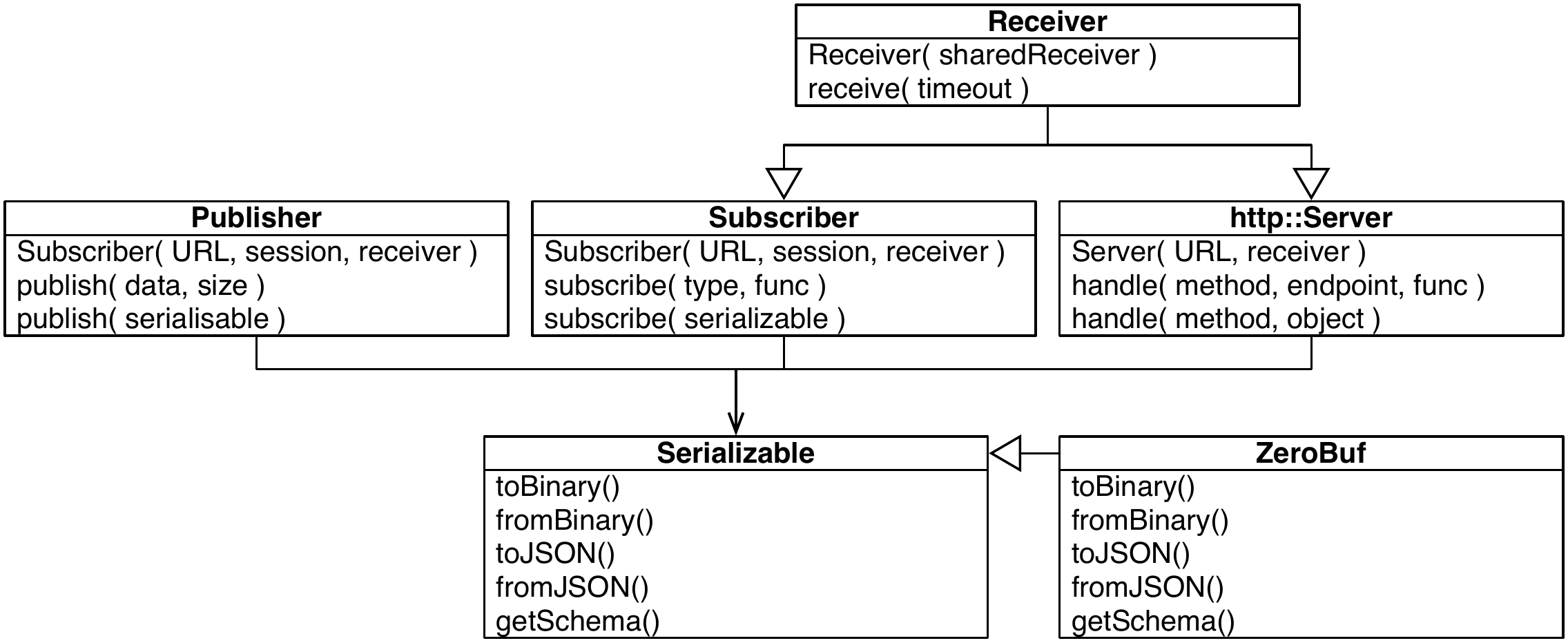}
  \caption{\label{fUML}UML Diagram of the main messaging classes}
\end{figure}

The base communication layer ZeroEQ utilizes ZeroMQ as the transport layer, the
ZeroConf protocol for discovery, and our novel ZeroBuf serialization library for
high-performance messaging. A fully integrated HTTP server provides a bridge to
JavaScript, Python and similar environments by implementing REST APIs with JSON
payload. \fig{fUML} shows a class diagram of our messaging solution.

\subsection{ZeroEQ}

ZeroEQ is our C++ messaging library, wrapping up existing technologies into an
API which is convenient to use and easy to integrate into C++ code. It provides
two messaging services: publish-subscribe and HTTP. For binary and JSON
encodings it relies on a simple \textsf{Serializable} interface, for which
ZeroBuf provides a sample implementation. To facilitate the simple use case of
linking a few applications, ZeroEQ uses the zeroconf protocol to discover and
connect to related applications. For more complex scenarios, explicit
connections are supported.

\subsubsection{Publish-Subscribe}

The publish-subscribe service is implemented in a \textsf{Publisher} and
\textsf{Subscriber} class. It provides event-based messaging, based on a 128-bit
message type with arbitrary payload. The message type is used for message
subscription, filtering and routing. The payload is expected to be uniquely
identified by the message type, that is, all applications agree for the decoding
and semantics of any given message type. ZeroBuf provides a sample
implementation for this. The underlying transport uses ZeroMQ pub-sub sockets.

The pub-sub service is stateless, that is, applications have no expectation of
when messages are received or who receives published messages. This
communication pattern naturally leads to robust services, since there is no
possiblity for deadlocks or undefined behaviour. The pub-sub API is provided in
two flavors: a simple $pointer\ \&\ size$ memory buffer, and a higher level
object-based abstraction. The object-based API is syntactic sugar for the
low-level API, and allows automatic publish and update of objects with a few
lines of code. It uses the \textsf{toBinary()} and \textsf{fromBinary()} methods
of the \textsf{Serializable} interface to call the low-level API.

\begin{lstlisting}[float, caption=Publish-Subscribe Example, label=lPubSub]
zerobuf::render::Camera camera;
zeroeq::Publisher publisher;
zeroeq::Subscriber subscriber;

subscriber.subscribe( camera );

while( rendering )
{
    if( updateCamera( camera )) // had user event
        publisher.publish( camera );
    else // poll subscription
        subscriber.receive( 0 /*ms*/ );
    renderFrame( camera );
}
\end{lstlisting}

The example in \lref{lPubSub} shows the integration of camera
synchronization in a visualization application. This example relies on the
builtin zeroconf protocol to connect application instances. Subscribers only
subscribe to events from publishers within the same session. The default
session name is the user name, and can be customized using an environment
variable or non-default constructor. Similarly, the subscriber can subscribe by
session or address. \lref{lPubSubExp} illustrates an explicitly addressed
subscription. Notice that the subscriber uses the publisher URI, which will
contain the concrete port chosen for the publisher.

\begin{lstlisting}[float, caption=Explicit Addressing,label=lPubSubExp]
zeroeq::URI uri( "tcp://localhost" );
zeroeq::Publisher publisher( uri, zeroeq::NULL_SESSION ); // deactivate zeroconf
zeroeq::Subscriber subscriber( publisher.getURI( )); // use concrete port
\end{lstlisting}

Subscribers are derived from a \textsf(Receiver) base class, which is shared
with the http server. All receivers can share their \textsf{receive()}
operation at construction time, that is, the blocking receive operation applies
to all receivers in the shared group. \lref{lSubShare} shows an example of
selectively receiving different updates on different input sockets.

\begin{lstlisting}[float,caption=Subscriber Sharing, label=lSubShare]
zeroeq::Subscriber local( zeroeq::URI( "localhost:29387" ));
zeroeq::Subscriber global( local );

local.subscribe( colorMap );
global.subscribe( camera );

while( true )
    local.receive(); // updates colorMap and camera
\end{lstlisting}

\subsubsection{HTTP Server}

The http server is built using cppnetlib~\cite{cppnetlib} for the transport and
http protocol handling. It supports all standard http verbs (GET, POST, PUT,
PATCH, DELETE). It is a \textsf{zeroeq::Receiver}, that is, it can share its
\textsf{receive()} update operation with other subscribers and http servers.
Unlike a subscriber, the http server follows the HTTP request-reply semantics,
that is, a request received by a server has to be followed directly by its
reply. To allow asynchronous request processing, the return value from the
request handler is a \textsf{std::future} which is retrieved from an internal
thread, thus allowing the application to continue operations.

The data served by the http server is introspectable, it allows querying the
available endpoints (objects) and the JSON schema~\cite{jsonschema} for each
endpoint. \lref{lhttpReg} shows an excerpt of the Tide registry, and
\lref{lhttpSchema} an excerpt of the schema for one of the exposed objects. This
REST API is used by the Tide web interface from Javascript and to generate
remote python APIs.

\noindent\begin{minipage}[b][][b]{.48\textwidth}
\begin{lstlisting}[caption=HTTP Server Registry, label=lhttpReg]
> GET /registry HTTP/1.0
{
[...]
   "tide/open" : [ "PUT" ],
   "tide/options" : [ "GET", "PUT" ],
   "tide/resize-window" : [ "PUT" ],
[...]
}
\end{lstlisting}
\end{minipage}\hfill
\begin{minipage}[b][][b]{.48\textwidth}
\begin{lstlisting}[caption=Object JSON Schema, label=lhttpSchema]
> GET /tide/options/schema HTTP/1.0
{
[...]
    "properties": {
        "alphaBlending": {
            "type": "boolean"
        },
[...]
}
\end{lstlisting}
\end{minipage}

\subsection{ZeroBuf}

ZeroBuf is a sample implementation of serialization for ZeroEQ. Based on a
grammar closely related to Flatbuffers schemas~\cite{flatbuffers}, it generates
C++ classes with random \textsf{set/get} access. All data is stored internally
in one continuous memory buffer, which can be used for the ZeroEQ binary
serialization without any copy. The conversion to and from a JSON representation
involves a copy using a \textsf{std::string}. ZeroBuf can store:

\begin{compactitem}
\item (u)int[8,16,32,64,128]\_t, float, double and string members
\item fixed size arrays and dynamic vectors of static-sized elements (intrinsic
members or composite types)
\item static and dynamic sub-classes (composite types of the above)
\end{compactitem}

\begin{figure}[ht]\center
  \includegraphics[width=\columnwidth]{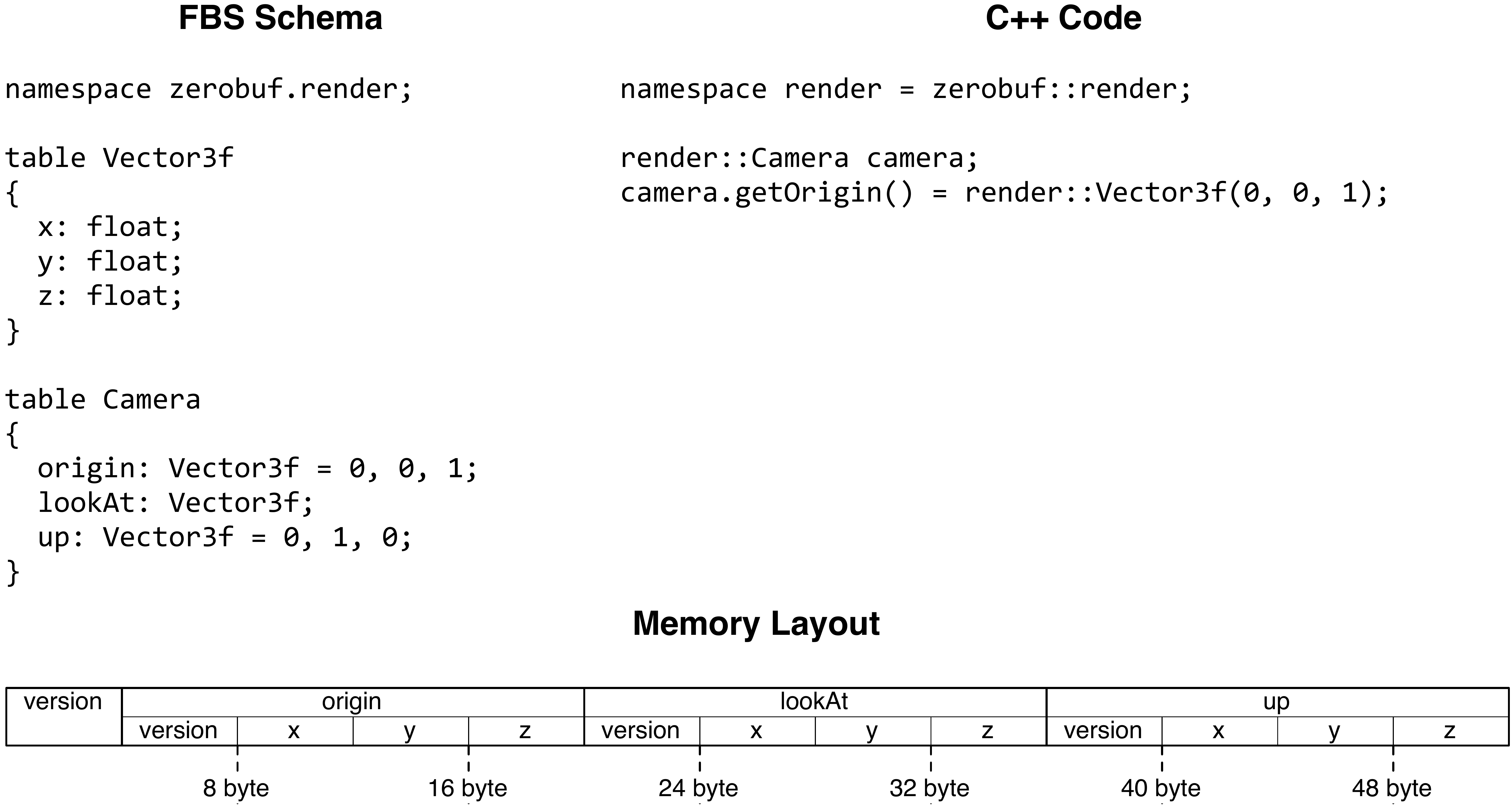}\\\vspace{4ex}
  \includegraphics[width=\columnwidth]{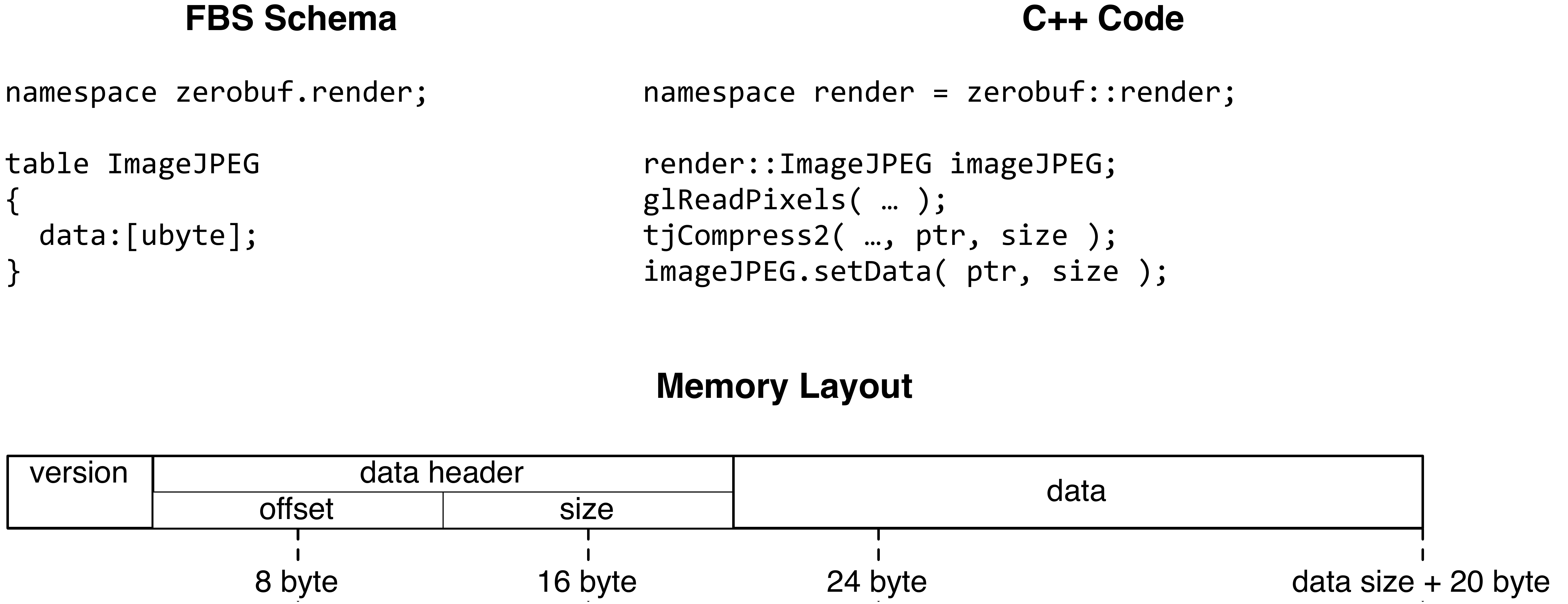}
  \caption{\label{fZeroBuf}ZeroBuf Examples for static (top) and dynamic
(bottom) sized objects}
\end{figure}

\fig{fZeroBuf} shows two simple ZeroBuf schemas together with example usage of
the generated code in C++ and their memory layout. The static example shows
nested ZeroBuf classes for the camera used in \lref{lPubSub}, and the dynamic
example shows how raw data access is used to prepare a JPEG image for
publishing.

\subsection{Remote Python API}

The remote python API provides easy to use access to remote applications using
the http server. It integrates two features: generic code generation for the
REST API exposed by the application, and automatic resource allocation and
application launch.

The generic code generation is implemented in a pure python module, which has
no dependency to the interfaced C++ application. It queries the http server and
generates a python API for all exposed objects. This API can then be
conveniently used in python to remote control the application.

Access to the application is established either through an explicit connection
of a pre-launched application, or via a resource allocator. The allocator hides
the details of allocating a resource, e.g. using a scheduling system like
slurm, launching and connecting to the launched application from the python
programmer.

\begin{figure}[ht]\center
  \includegraphics[width=\columnwidth]{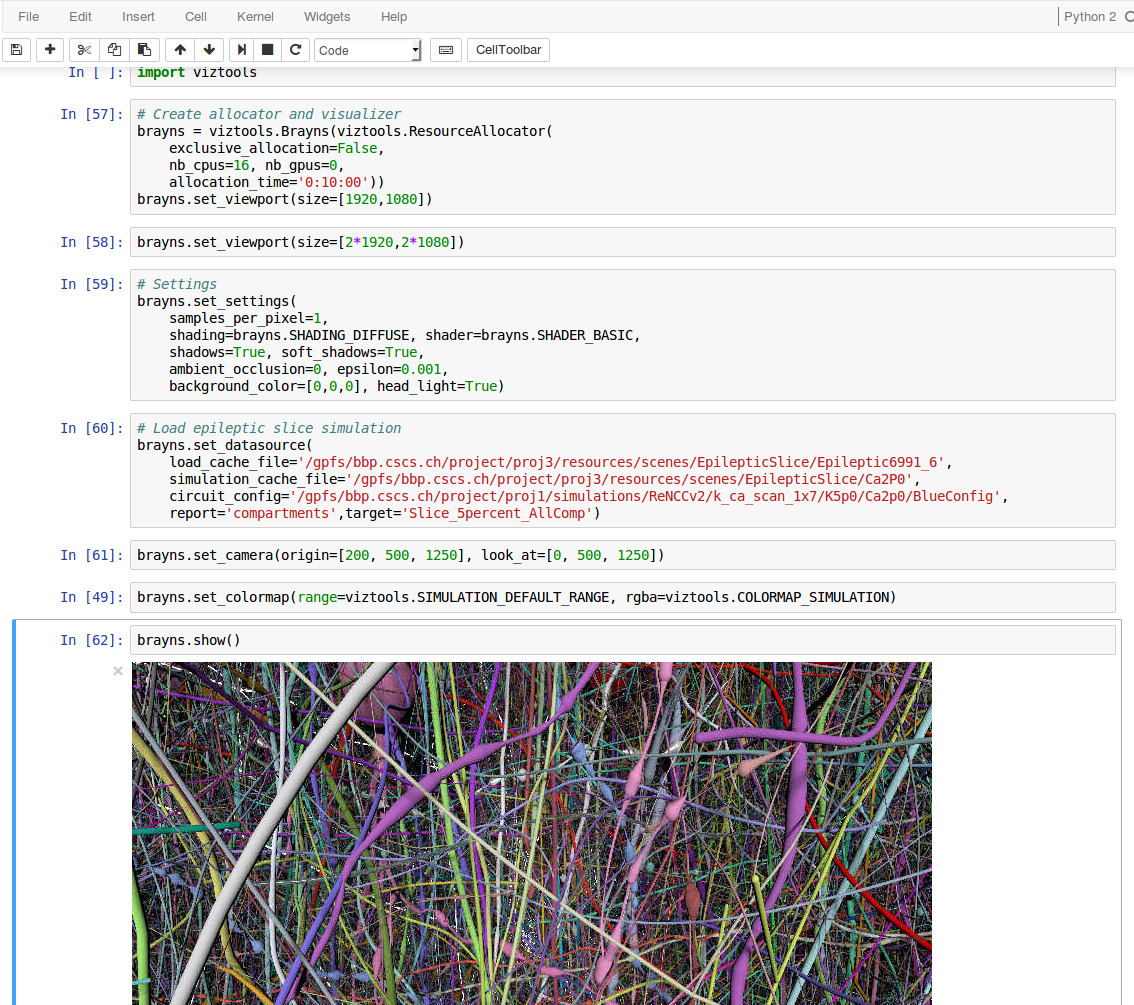}
  \caption{\label{fJupyter}Example Jupyter notebook session using Brayns}
\end{figure}

\fig{fJupyter} shows an example session of using this Python API from a Jupyter
notebook, allocating and launching a Brayns instance, setting relevant rendering
parameters and retrieving an image. Note that the whole notebook runs in a
light-weight VM with no GPU and interacts with a Brayns instance launched on a
bare-metal visualization cluster node.

\section{Discussion and Conclusion}
\label{sec:conclusions}

We presented a modular visualization architecture for large data visualization
over a wide range of use cases, glued together by a modern and easy to use
messaging infrastructure. This ecosystem allows us to flexibly support novel use
cases, while pushing novel visualization capabilities, providing Blue Brain with
a competitive visualization infrastructure. Messaging and remote APIs not only
surprised us in their versatility and ease of integration with other ecosystems,
but also have a significant potential for future exploration in classical
visualization software. Interactive raytracing is the future rendering algorithm
for us, and Brayns is becoming the integration point for our domain-specific
visual applications and algorithms. Tiled display walls are affordable for a
large set of institutions, and coupled with our open source TIDE software create
new ways of truely collaborative work. TIDE, together with cheaper visualization
hardware, will evolve in the future towards a seamless integration of immersive
visualization.

\section*{Acknowledgments}

This publication was supported by the Blue Brain Project (BBP), the Swiss
National Science Foundation under Grant 200020-129525, the King Abdullah
University of Science and Technology (KAUST) through the KAUST-EPFL alliance for
Neuro-Inspired High Performance Computing, the Spanish Ministry of Science and
Innovation under grant (TIN2010-21289-C02-01/02), the Cajal Blue Brain Project,
Hasler Stiftung Projekt Nr. 12097, and from the European Union’s Horizon 2020
research and innovation programme under grant agreement No 720270 (HBP SGA1). We
would also like to thank the people from GMRV at the Rey Juan Carlos University
(URJC) for their collaboration under the Cajal Blue Brain and HBP projects.

\bibliographystyle{abbrv}
\bibliography{references/references}

\begin{thebibliography}{10}

\bibitem{tide}
{Blue Brain Project}.
\newblock {Tide: Tiled Interactive Display Environment}.
\newblock https://github.com/BlueBrain/Tide, 2016.

\bibitem{brayns}
{Blue Brain Project}.
\newblock {Brayns: Interactive raytracing of neuroscience data}.
\newblock https://github.com/BlueBrain/Brayns, 2017.

\bibitem{Sage}
T.~A. DeFanti, J.~Leigh, L.~Renambot, B.~Jeong, A.~Verlo, L.~Long, M.~Brown,
  D.~J. Sandin, V.~Vishwanath, Q.~Liu, M.~J. Katz, P.~Papadopoulos, J.~P.
  Keefe, G.~R. Hidley, G.~L. Dawe, I.~Kaufman, B.~Glogowski, K.-U. Doerr,
  R.~Singh, J.~Girado, J.~P. Schulze, F.~Kuester, and L.~Smarr.
\newblock The optiportal, a scalable visualization, storage, and computing
  interface device for the optiputer.
\newblock {\em Future Gener. Comput. Syst.}, 25(2):114--123, Feb. 2009.

\bibitem{DK:11}
K.-U. Doerr and F.~Kuester.
\newblock {CGLX}: A scalable, high-performance visualization framework for
  networked display environments.
\newblock {\em IEEE Transactions on Visualization and Computer Graphics},
  17(2):320--332, March 2011.

\bibitem{EMP:09}
S.~Eilemann, M.~Makhinya, and R.~Pajarola.
\newblock Equalizer: A scalable parallel rendering framework.
\newblock {\em IEEE Transactions on Visualization and Computer Graphics},
  15(3):436--452, May/June 2009.

\bibitem{Omegalib}
A.~Febretti, A.~Nishimoto, V.~Mateevitsi, L.~Renambot, A.~Johnson, and
  J.~Leigh.
\newblock {Omegalib: A multi-view application framework for hybrid reality
  display environments}.
\newblock In {\em 2014 IEEE Virtual Reality (VR)}, pages 9--14, March 2014.

\bibitem{flatbuffers}
{Google, Inc}.
\newblock {Cross Platform Serialization Library}.
\newblock http://google.github.io/flatbuffers/, 2017.

\bibitem{cppnetlib}
D.~M. B. G.~M. {Google, Inc}.
\newblock {The C++ Network Library Project}.
\newblock http://cpp-netlib.org/, 2017.

\bibitem{guthe2004}
S.~Guthe and W.~Strasser.
\newblock Advanced techniques for high-quality multi-resolution volume
  rendering.
\newblock {\em Computers \& Graphics}, 28(1):51--58, 2004.

\bibitem{HBBES:13}
J.~B. Hernando, J.~Biddiscombe, B.~Bohara, S.~Eilemann, and F.~Sch\"{u}rmann.
\newblock {Practical Parallel Rendering of Detailed Neuron Simulations}.
\newblock In {\em Proceedings of the 13th Eurographics Symposium on Parallel
  Graphics and Visualization}, EGPGV, pages 49--56, Aire-la-Ville, Switzerland,
  Switzerland, 2013. Eurographics Association.

\bibitem{JLMV:06}
A.~Johnson, J.~Leigh, P.~Morin, and P.~Van~Keken.
\newblock {GeoWall}: Stereoscopic visualization for geoscience research and
  education.
\newblock {\em IEEE Computer Graphics and Applications}, 26(6):10--14,
  November-December 2006.

\bibitem{DisplayCluster}
G.~P. Johnson, G.~D. Abram, B.~Westing, P.~Navr'til, and K.~Gaither.
\newblock {DisplayCluster: An Interactive Visualization Environment for Tiled
  Displays}.
\newblock In {\em 2012 IEEE International Conference on Cluster Computing},
  pages 239--247, Sept 2012.

\bibitem{jsonschema}
{JSON Schema}.
\newblock {JSON Schema}.
\newblock http://json-schema.org/, 2017.

\bibitem{Sage2}
T.~Marrinan, J.~Aurisano, A.~Nishimoto, K.~Bharadwaj, V.~Mateevitsi,
  L.~Renambot, L.~Long, A.~Johnson, and J.~Leigh.
\newblock {SAGE2: A new approach for data intensive collaboration using
  Scalable Resolution Shared Displays}.
\newblock In {\em Collaborative Computing: Networking, Applications and
  Worksharing}, pages 177--186, 2014.

\bibitem{optix}
S.~G. Parker, J.~Bigler, A.~Dietrich, H.~Friedrich, J.~Hoberock, D.~Luebke,
  D.~McAllister, M.~McGuire, K.~Morley, A.~Robison, and M.~Stich.
\newblock {OptiX: A General Purpose Ray Tracing Engine}.
\newblock {\em ACM Transactions on Graphics}, August 2010.

\bibitem{ospray}
I.~Wald, G.~Johnson, J.~Amstutz, C.~Brownlee, A.~Knoll, J.~Jeffers,
  J.~Günther, and P.~Navratil.
\newblock {OSPRay - A CPU Ray Tracing Framework for Scientific Visualization}.
\newblock {\em IEEE Transactions on Visualization and Computer Graphics},
  23(1):931--940, Jan 2017.

\end{thebibliography}

\end{document}